\documentclass[review]{elsarticle}

\usepackage{graphicx}
\usepackage{dcolumn}
\usepackage{pifont}
\usepackage{bm}
\usepackage{multirow}
\usepackage{amsmath}
\usepackage{float}
\usepackage{txfonts}
\usepackage[version=3]{mhchem}

\def\mathbi#1{\textbf{\em #1}}

\usepackage[usenames,dvipsnames]{xcolor}
\definecolor{cream}{RGB}{222,217,201}
\definecolor{red}{RGB}{225,0,0}

\journal{Cement and Concrete Research}

\begin{document}
\title{Thermal conductivity of porous jennite by molecular dynamics method}

\author[kimuniv-m]{Song-Nam Hong}
\author[kimuniv-m]{Chol-Jun Yu\corref{cor}}
\cortext[cor]{Corresponding author}
\ead{cj.yu@ryongnamsan.edu.kp}
\author[kimuniv-m]{Un-Song Hwang}
\author[kimuniv-m]{Chung-Hyok Kim}
\author[kumsu]{Byong-Hyok Ri}

\address[kimuniv-m]{Chair of Computational Materials Design (CMD), Faculty of Materials Science, Kim Il Sung University, Ryongnam-Dong, Taesong District, Pyongyang, Democratic People's Republic of Korea}
\address[kumsu]{Changhun-Kumsu Agency for New Technology Exchange, Mangyongdae District, Pyongyang, Democratic People's Republic of Korea}

\begin{abstract}
The thermal conductivity of porous jennite, as the major component of cement paste, and its porosity and temperature dependences are simulated by molecular dynamics methods using ClayFF force field.
The porous jennite models with different porosities are created by removing atoms within the sphere from bulk jennite model.
The thermal conductivity elements of bulk jennite calculated by equilibrium Green-Kubo method are almost identical to those by non-equilibrium M\"{u}ller-Plathe (MP) method at 300 K.
The volumetric thermal conductivity of porous jennite is found to decrease from 1.141 W/m$\cdot$K to 0.144 W/m$\cdot$K as increasing the porosity from 0\% to 72.22\% at 300 K, following the empirical coherent potential model when the pore is assumed to be filled with air.
As increasing temperature, the thermal conductivity is observed to increase from 240 K to 560 K and gradually decrease until 1100 K for porous jennites with the porosities of 0\%, 15.99\% and 32.79\%.
\end{abstract}


\maketitle

\section{\label{sec:intro}Introduction}
Building materials have always been indispensable for survival of human beings, and among them, concrete has been the most widely used since the invention of modern cement.
While concrete consists of cement matrix and aggregates such as sand and gravel or industrial slag, its main binding phase is known to be calcium silicate hydrates (CaO$-$\ce{SiO2}$-$\ce{H2O}; C$-$S$-$H) gel produced by cement hydration~\cite{Richardson07ccr,Gastaldi12cbm}.
The C$-$S$-$H gel is responsible for various important properties of concrete, including strength, toughness, and durability.
Therefore, considerably mass of experimental and theoretical research has been carried out in revealing and improving mechanical properties of concrete and C$-$S$-$H gels~\cite{Fan, Al-Ostaz, Zaoui, Hajilar, Skinner, Masoero}.

Although atomic modeling is widely accepted as a powerful tool to provide invaluable insights into the fundamental behavior of materials, the modeling of real C$-$S$-$H gel is known to be exceedingly challenging.
This is partly associated with the difficulties in identifying the atomistic structures of C$-$S$-$H gel by material characterization experiments, such as X-ray diffraction (XRD), transition or scanning electron microscopy (TEM or SEM) and Raman spectroscopy~\cite{Richardson}.
In fact, the C$-$S$-$H gel in concrete is imperfect crystalline, in which the calcium to silicon (C/S) ratio is changeable from 1.2 to 2.1 according to the hydration process condition~\cite{Richardson99ccr}.
In such circumstances, they are often plausibly modeled by tobermorite and jennite minerals based on their structural similarities observed by XRD~\cite{Taylor05jacs}, although more realistic model with a C/S ratio of 1.65 for C$-$S$-$H has been suggested by randomly removing charge-neutralized \ce{SiO2} units from tobermorite 11 \AA~\cite{Pellenq,Allen,Hou, Hou2, Qomi}. 
The mineral tobermorite has much smaller C/S ratio of $\sim$0.83 compared with experiment, whereas for jennite it is reasonable to be about 1.5 due to the corrugation of Ca$-$O sheets~\cite{Vidmer14ccr}.
So far, jennite and tobermorite family have been intensively studied as substitutes for the C-S-H gels.
Bonaccorsi {\it et al}. have done pioneering work on characterizing the crystalline structures of jennite~\cite{Bonaccorsi-jen}, tobermorite 11 \AA~\cite{Merlino01ejm} and tobermorite 14 \AA~\cite{Bonaccorsi-tob} through XRD experiments.
The structural and mechanical properties of jennite and tobermorite have been investigated using first-principles calculations,~\cite{Vidmer14ccr, Dharmawardhana, Churakov08ccr, Shahsavari09jacs, Moon, Tunega, Rejmak} and molecular dynamics (MD) simulations~\cite{Dongshuai, Manzano, Mutisya, Hou1}.
However, there is only a limited number of studies on thermal properties of jennite and tobermorite as well as C$-$S$-$H gel~\cite{Yoon, Xu, Bentz, Qomi1, Misri}, in spite of its importance in design and construction of energy efficient buildings with concrete~\cite{Sabnisbook12}.
In particular, studies of revealing the thermal conductivity of jennite at molecular scale are truly scarce.

Recently, there is a growing interest in high thermal insulation materials in construction to save energy use in buildings~\cite{Gobakis15eb, Papadopoulos}.
In general, thermal insulation materials are porous~\cite{Jambor90ccr, Bhutta}.
The thermal conductivity of such porous materials decreases as their porosity increases due to an increasing amount of inside small-sized air pores.
Porous concrete is in particular advantageous due to its high thermal insulation, low density, fire-proofing and acoustic insulation~\cite{Hatanaka}.
Qomi {\it et al}.~\cite{Qomi1} reported the thermal properties of C$-$S$-$H, such as specific-heat capacity, thermal expansion coefficient and thermal conductivity, by means of MD method using core-only CSH force field potential~\cite{Shahsavari}. 
Jin {\it et al}.~\cite{QingJin} proposed the fractal model for predicting the thermal conductivity of porous concrete based on their measurements of autoclaved aerated concrete with various porosities.

In this work, we aim to determine the thermal conductivity of jennite and its dependence on the porosity and temperature by using MD simulations.
We organized the paper as follows.
In Section~\ref{sec_method}, we described computational methods, including the employed force field potential, jennite models and MD methods for predicting the thermal conductivity.
Section~\ref{sec_result} shows the results and discussion, where comparisons with the previous MD works, experimental results, and predictions based on the continuum models for heat transfer in porous materials, were provided.
Temperature dependence of thermal conductivity was also presented together with a discussion in comparison with the available literature data.
Finally, we provided the conclusions of this work in Section~\ref{sec_con}.

\section{\label{sec_method}Methods}
\subsection{\label{sub_force}Force field}
The accuracy of classical MD simulations strongly depends on the selected interatomic force field potential.
To date, many force fields have been developed for cementitious materials~\cite{Mishra, Lau}.
They are for example ClayFF~\cite{Cygan}, IFF (interface force field)~\cite{Heinz}, CementFF~\cite{Freeman}, ReaxFF~\cite{Duin}, and CSH-FF~\cite{Shahsavari}.
Among them, the ClayFF potential was selected in this work, since it has been completely tested through lots of studies on various properties of clays and cement materials~\cite{Hou1, Kalinichev, Kalinichev1, Tavakoli}.
As Mutisya {\it et al}.~\cite{Mutisya} pointed out, moreover, the ClayFF is able to capture thermodynamic properties of both hydrous and anhydrous cement phases.

In the ClayFF formalism, the total energy of a molecular system is represented as follows~\cite{Mishra, Cygan},
\begin{equation}
\label{eq_clayff_total_energy}
E_{\text{total}}=E_{\text{Coul}}+E_{\text{vdW}}+E_{\text{bond}}+E_{\text{angle}}
\end{equation}
where $E_{\text{Coul}}$ is the Coulomb interaction energy, $E_{\text{vdW}}$ is the short-range van der Waals (vdW) energy, $E_{\text{bond}}$ is the bond stretching term, and $E_{\text{angle}}$ is the angle bending term, respectively.
The electrostatic interaction energy between the charged particles is described by the Coulomb's law as,
\begin{equation}
\label{eq_coulmb}
E_{\text{Coul}}=\frac{e^2}{4\pi\varepsilon_0}\sum_{i\neq j}\frac{q_iq_j}{r_{ij}}
\end{equation}
where $q_i$ is the atomic partial charge of $i$th particle, $r_{ij}$ is the distance between the $i$th and $j$th particles, $e$ is the elementary charge, and $\varepsilon_0$ is the dielectric permittivity of vacuum.
The attractive vdW dispersion energy is written as~\cite{Mishra, Cygan},
\begin{equation}
\label{eq_vdw}
E_{\text{vdW}}=\sum_{i\neq j} D^0_{ij} \left[\left(\frac{R^0_{ij}}{r_{ij}}\right)^{12} - 2\left(\frac{R^0_{ij}}{r_{ij}}\right)^6\right]
\end{equation}
where $D^0_{ij}=\sqrt{D^0_iD^0_j}$ and $R^0_{ij}=(R^0_i+R^0_j)/2$. $D^0_{i}$, $R^0_{i}$ are the empirical parameters derived from the experimental data for physical property.
The bonded interactions, which are used to describe the O$-$H bonding in water molecules and hydroxyl groups, and the covalent bonding in polyatomic species, are simplified as harmonic terms as follows~\cite{Mishra, Cygan},
\begin{eqnarray}
\label{eq_bond}
E_{\text{bond}}=\sum_{i\neq j} k_1(r_{ij}-r_0)^2 \\
E_{\text{angle}}=\sum_{i\neq j\neq k} k_2(\theta_{ijk}-\theta_0)^2
\end{eqnarray}
where $k_1$ and $k_2$ are the harmonic force constant parameters, and $r_0$ and $\theta_0$ are the equilibrium values of bond length and bond angle respectively.
The parameters $q_i$, $D^0_i$, $R^0_i$ and $k$ for Ca, Si, O and H atoms considered in this work were generated by using the $msi2lmp$ tool of Large Scale Atomic/Molecular Massively Parallel Simulator (LAMMPS)~\cite{lammps}, and given in Table S1$-$S3 (Supplementary data).

\subsection{\label{sub_model}Models for porous jennite}
\begin{figure*}[!th]
\centering
\includegraphics[clip=true,scale=0.15]{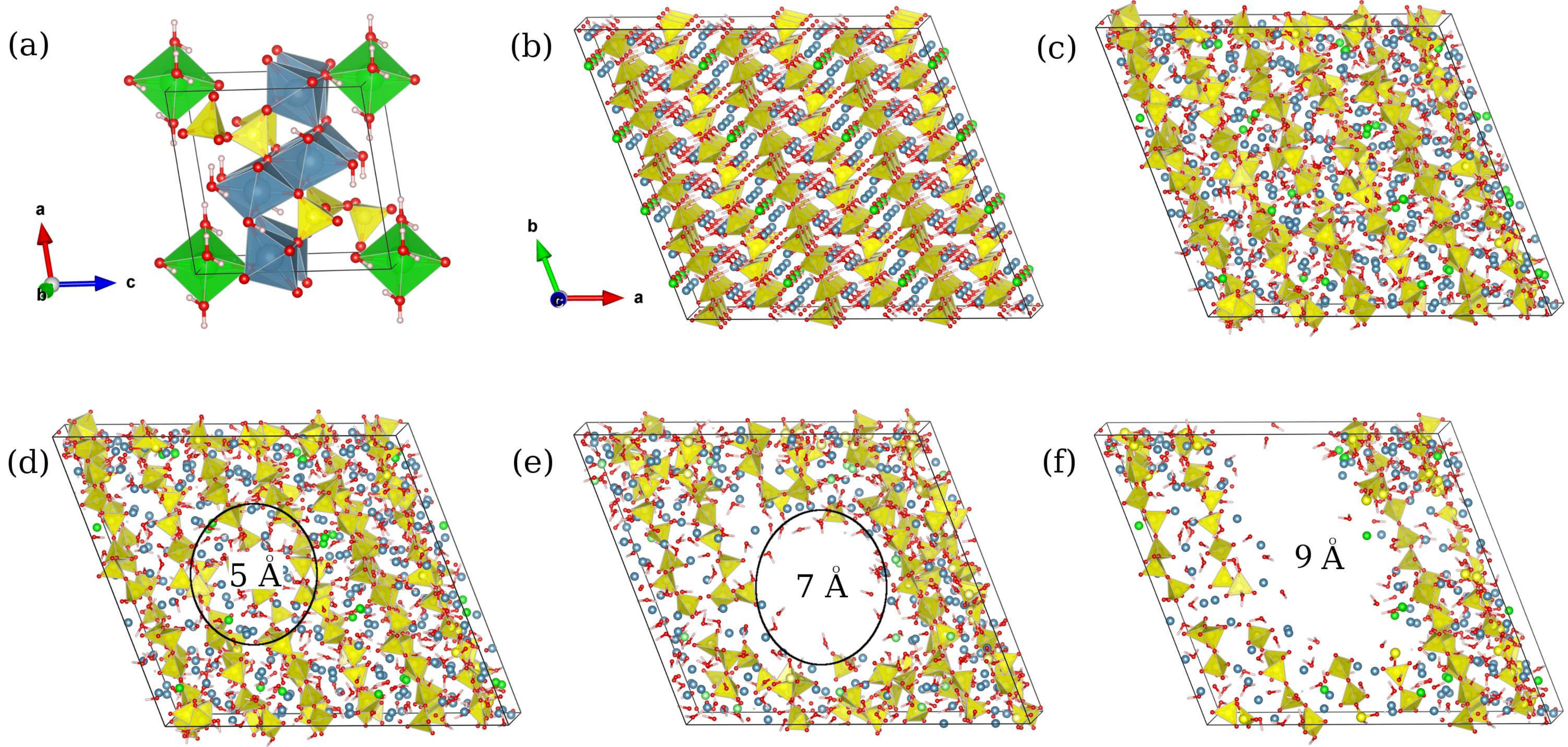} 
\caption{\label{fig_jennite}Polyhedral view of atomistic structures of jennite models. (a) Unit cell of jennite crystal, b) supercell for bulk jennite model with $3\times4\times3$ unit cells and c) its equilibrated phase, and porous jennite models created by removing atoms within sphere with radii of (d) 5 \AA, (e) 7 \AA~and (f) 9 \AA. Blue, yellow, red, white, and green balls indicate Ca, Si, O, H, and interlayer Ca atoms, and yellow polyhedra represent the \ce{SiO4}.}
\end{figure*}
The chemical formula of jennite is \ce{Ca9Si6O_{18}(OH)6\cdot}8\ce{H2O}, and its crystalline lattice is triclinic with $P$\={1} space group~\cite{Vidmer14ccr, Bonaccorsi-jen}.
Fig.~\ref{fig_jennite}(a) shows the unit cell of jennite (see jennite-unitcell.cif in Supplementary material).
For MD simulations, we constructed a supercell for bulk jennite model by using $3\times 4\times 3$ unit cells, which contains 2484 atoms, and subsequently carried out constant pressure and temperature ($NPT$) simulation with a time step of 0.1 fs for 100 ps at 300 K and atmospheric pressure, then it's size was about $32.6\times29.6\times32.8$ \AA, as shown in Fig.~\ref{fig_jennite}(b).
At this NPT simulation the cutoff radii for the LJ and Coulombic interactions were set to be 15 \AA.
The periodic boundary condition was applied in all three crystallographic directions, and the $pppm$ method was used for calculation of electrostatic term.
Nose-Hoover thermostat and barostat were used to control temperature and pressure of the system, and the velocity Verlet time integration algorithm was used to update the position and velocity of atoms.
Fig.~\ref{fig_jennite}(c) shows the equilibrated atomistic structure of the jennite model.
Table~\ref{tbl_jennite_structure} presents the determined lattice parameters, which are in good agreement with experiment with a maximal relative error of 2.73\%~\cite{Bonaccorsi-jen}.
\begin{table}[!b]
\small
\caption{\label{tbl_jennite_structure}Lattice parameters of unit cell of jennite crystal, equilibrated through $NPT$ run, in comparison with experiment.}
\begin{tabular}{lcccccc}
\hline
&$a$ (\AA)&$b$ (\AA)&$c$ (\AA)&$\alpha$ ($^\circ$)&$ \beta$ ($^\circ$)&$\gamma$ ($^\circ$)\\
\hline
ClayFF        & 10.865 & 7.402  & 10.946  & 102.06 & 94.57 & 110.33 \\
Exp~\cite{Bonaccorsi-jen} & 10.576 & 7.265  & 10.931  & 101.30   & 96.98  & 109.65\\
\hline
\end{tabular} 
\end{table}

To verify the validity of jennite model and the force field, we also evaluated the radial distribution function (RDF) of Ca$-$O in comparison with available experiment and previous MD simulations.
According to the difference of local structures, calcium atoms are classified into sheet (Ca$_{\text{s}}$) and interlayer (Ca$_{\text{w}}$) atoms, while oxygen atoms are divided into surrounding (O$_{\text{s}}$) and water (O$_{\text{w}}$) oxygen atoms.
Therefore, we considered four different RDFs for Ca$-$O bonds (see Fig. S1).
For the Ca$_{\text{s}}-$O bonds, the first peaks were found at 2.41 and 2.59 \AA, which are slightly larger than the previous MD results of 2.37 and 2.57 \AA~\cite{Dongshuai} while in reasonable agreement with the experimental value of 2.45$\pm$0.06 \AA~\cite{Bowers}.
The distances between Ca$_{\text{w}}$ and O$_{\text{w}}$ or O$_{\text{h}}$ (hydroxyl groups) atoms were estimated to be 2.35 or 2.50 \AA, being in good agreement with the XRD experimental data of 2.36 or 2.50 \AA~\cite{Bonaccorsi-jen}.
These indicate that our jennite model can be used safely and the ClayFF employed in this work is reasonably accurate and transferable for jennite.

Then, the porous jennite models were created by removing all the atoms within sphere with a certain radius located at the center of perfect jennite model.
Note that if one atom of a polyatomic molecule (e.g., \ce{CaO}, \ce{SiO2}, \ce{H2O}) belonged to the spherical region, the molecule was totally removed.
We increased the radius gradually from 3 \AA~to 11 \AA, and estimated the corresponding mass density and porosity of each porous model.
The porosity can be estimated by using the mass densities of perfect model $\rho_0$ and porous model $\rho_p$ as follows,
\begin{equation}
\label{eq_porosity}
P=\left(1-\frac{\rho_p}{\rho_0}\right)\times 100\%
\end{equation}
For reference, we also provided the porosity calculated using the number of atoms instead of density.
Table~\ref{tbl_model} lists the characteristics of porous jennite models with various porosities from 0 to 72.22\%.
Fig.~\ref{fig_jennite}(d), (e) and (f) show the equilibrated atomistic structures of porous jennite models with the porosities of 10.39\%, 32.79\% and 52.39\% respectively.
\begin{table}[!t]
\small
\caption{\label{tbl_model} Sphere radius ($r$), total number of atoms ($N_{\text{tot}}$), number of deleted atoms ($N_{\text{del}}$), mass density ($\rho$), and porosities estimated using number of atoms ($P_N$) and mass density ($P_{\rho}$) in perfect and porous jennite models.}
\begin{tabular}{ccccccc}
\hline
model & $r$ (\AA) & $N_{\text{tot}}$ & $N_{\text{del}}$ & $\rho$ (g/cm$^3$) & $P_N$ (\%) & $P_{\rho}$ (\%) \\
\hline
1 & 0   & 2484 & 0    & 2.205 & 0     & 0 \\
2 & 3   & 2346 & 138  & 2.099 & 5.56  & 4.79 \\
3 & 5   & 2208 & 276  & 1.976 & 11.11 & 10.39 \\
4 & 5.5 & 2070 & 414  & 1.852 & 16.67 & 15.99 \\
5 & 6   & 1863 & 621  & 1.667 & 25.00 & 24.39 \\
6 & 7   & 1656 & 828  & 1.482 & 33.33 & 32.79 \\
7 & 9   & 1173 & 1311 & 1.050 & 52.78 & 52.39 \\
8 & 11  & 690  & 1794 & 0.612 & 72.22 & 72.22 \\
\hline
\end{tabular} 
\end{table}

\subsection{\label{sub_process}Molecular dynamics methods for predicting the thermal conductivity}
The thermal conductivity $\mathbi{K}$ of insulating solid is defined as a second-order tensor coefficient relating the temperature gradient $\nabla T$ to the heat flux $\mathbi{q}$ through Fourier's law, $\mathbi{q}=\mathbi{K}\cdot \nabla T$. 
Typically, two different methods have been developed to calculate $\mathbi{K}$ via MD simulation: (1) equilibrium Green-Kubo (GK) method~\cite{Green, Kubo} and (2) non-equilibrium M\"{u}ller-Plathe (MP) method~\cite{Muller}.
The former method does not impose temperature gradients and yields the full thermal conductivity tensor via one time simulation, while the latter approach imposes a heat flux on the simulation box and gives the heat conductivity separately in each direction for an anisotropic material.
In this work, we applied these two approaches in order to guarantee an accuracy of calculation, relying mostly on the MP method.

In the GK approach, once the molecular system is equilibrated through $NPT$ simulation, the net heat flux fluctuates around zero during $NVE$ run.
The dissipation rate of heat flux fluctuations is evaluated, and the thermal conductivity tensor for an anisotropic materials is measured at equilibrium by using the following GK relation~\cite{Qomi1, Frenkel},
\begin{equation}
\label{eq_green_kubo}
K_{\alpha\beta}=\frac{V}{k_BT^2} \int_{0}^{\infty}\left\langle q_{\alpha}(t)\otimes q_{\beta}(0)\right\rangle dt 
\end{equation}
where $\otimes$ is the dyadic product in tensor notation, $k_B$ the Boltzmann constant, $V$ the volume of system, and $\alpha$, $\beta$ the components of coordinate system.
In this equation, $\phi_{\alpha\beta}(t)=\langle q_{\alpha}(t) \otimes q_{\beta}(0) \rangle / \langle q_{\alpha}(0) \otimes q_{\beta}(0)\rangle$ is the second-order tensor known as the heat flux autocorrelation function (HFACF), and thus, the thermal conductivity can be calculated from integration of HFACF. 
The heat flux in a polyatomic system with a number of atom $N$ is written as follows~\cite{Qomi1},
\begin{equation}
\label{eq_heat_flux}
q_{\alpha}(t)=\frac{1}{V}\frac{d}{dt}\sum_{i=1}^{N}r_{i,\alpha}H_i 
\end{equation}
%
%
where $\mathbi{r}_i$ is the position vector of the $i$th atom, and $H_i$ is the Hamiltonian, which is composed of kinetic and potential (Eq.~(\ref{eq_clayff_total_energy})) energies.
%
The GK approach has been applied to C$-$S$-$H~\cite{Qomi1} and amorphous silicon dioxide thin films~\cite{Gu}.
In this work, the equilibrium GK approach was applied to the perfect jennite crystal with the $3\times 4\times 3$ supercell using LAMMPS~\cite{lammps}.
After equilibration of the system through $NPT$ run, we subsequently performed $NVE$ simulation with a time step of 0.1 fs for 20 ns ($2\times10^5$ correlation time steps).

In the non-equilibrium MP method, thermal conductivity can be calculated by imposing a heat flux on the simulation box and measuring the resultant temperature gradient~\cite{Muller, Mahajan07pre, Coquil, Zhu}.
If the heat propagates along $\alpha$ direction in the simulation box, the thermal conductivity in this direction is calculated by
\begin{equation}
\label{eq_mp}
K_{\alpha}=-\frac{q_{\alpha}}{\partial T / \partial r_{\alpha}}=-\frac{q_{\alpha}}{\Delta T / \Delta r_{\alpha}}
\end{equation}
As implemented in the LAMMPS~\cite{lammps}, the simulation box is divided into an even number of parallel slices, and the temperature of each slice is evaluated at every time step and then averaged over the whole time steps~\cite{Coquil}.
The heat flux is imposed through velocity rescaling, i.e., by exchanging velocities between atoms with the largest kinetic energy of the first slice and atoms with the lowest kinetic energy of the middle slice at a certain simulation time step.
After reaching steady state by imposing heat flux for sufficiently long time, a temperature profile $T(r_{\alpha})$ is evaluated and thus the thermal conductivity is ready for estimation by using Eq.~(\ref{eq_mp}).
For an anisotropic material, this procedure has to be performed separately to obtain the thermal conductivity in each direction of coordinate system.
The MP method is easy to understand and has moderate simulation time, so it has been used in many studies of porous materials~\cite{Gu, Coquil, Zhu}.

In this work, we applied the MP method mostly to the porous jennite models with various porosities for estimating their thermal conductivities at different temperatures, using the LAMMPS program~\cite{lammps}.
The porous jennite models, constructed by removing atoms within the spherical region as described above, were equilibrated by performing constant volume and constant temperature ($NVT$) simulations with a time step of 0.1 fs for 100 ps.
Then, the simulation boxes were created by extending these equilibrated porous jennite models by from two to ten times in [100], [010] and [001] crystallographic directions, and also equilibrated by $NVT$ run.
To calibrate the temperature profile and the thermal conductivity, the extended simulations boxes are divided into 20 plane-parallel slices, and micro canonical ensemble ($NVE$) simulations were performed with a time step of 0.1 fs during 200 ps.
Like the application of GK approach, the Nose-Hoover thermostat and barostat, and velocity Verlet time integration algorithm were adopted.

\section{\label{sec_result}Results and discussion}
\subsection{\label{sub_gk}Thermal conductivity of jennite by Green-Kubo method}
\begin{figure}[!b]
\centering
\includegraphics[clip=true,scale=0.5]{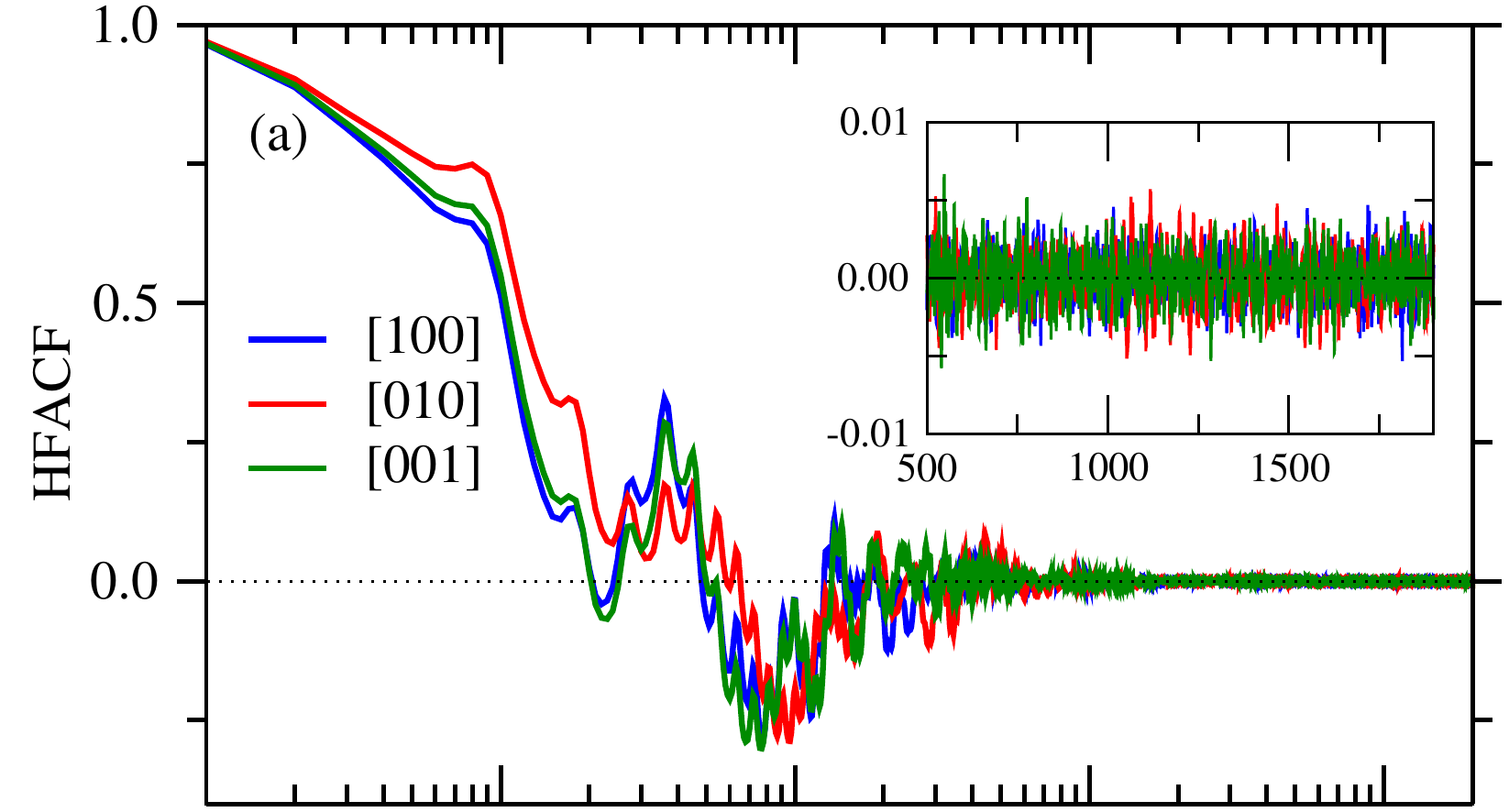}
\includegraphics[clip=true,scale=0.5]{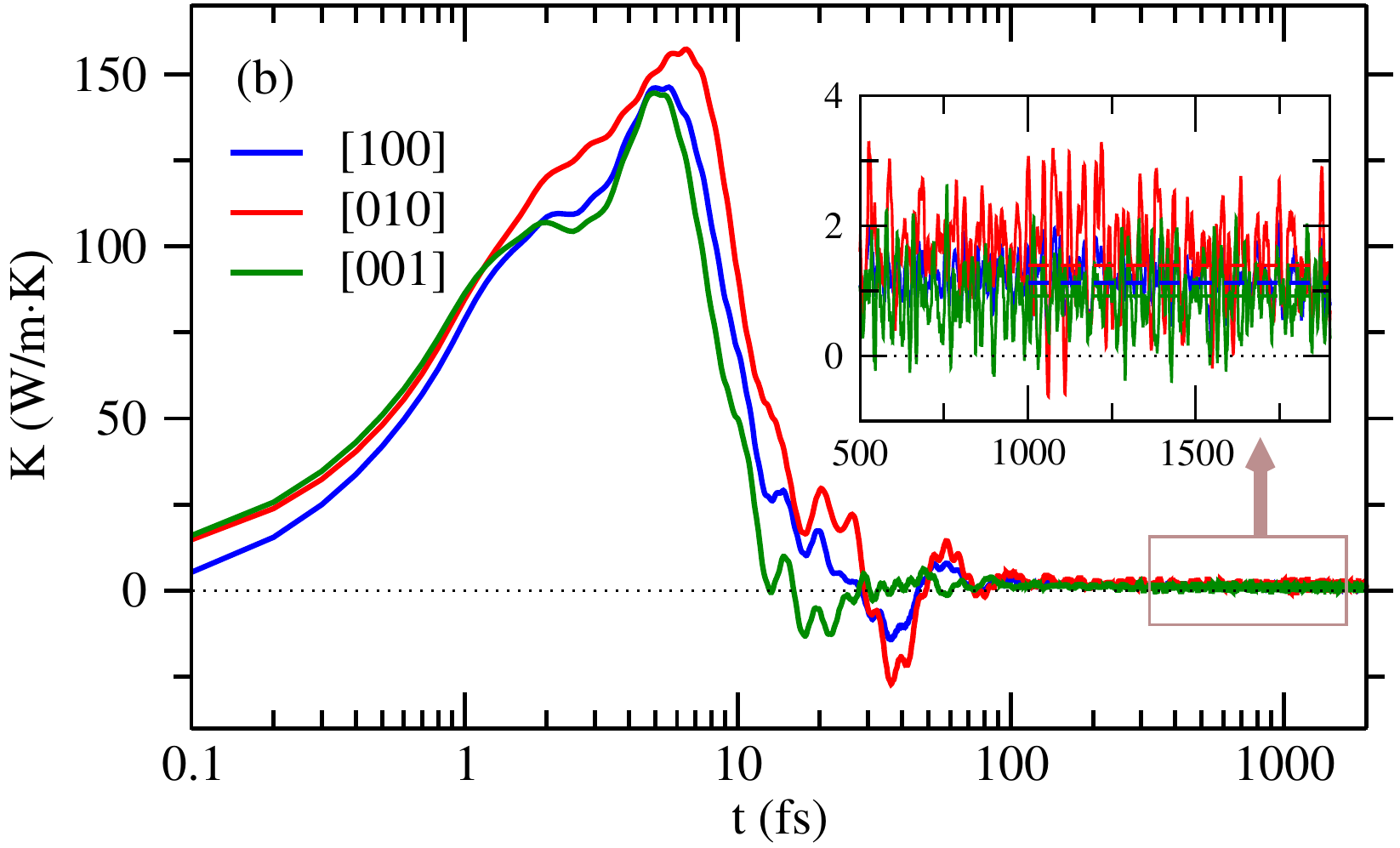}
\caption{\label{fig_green_kubo}(a) The heat flux autocorrelation function (HFACF) and (b) the thermal conductivity along the crystallographic directions as functions of correlation time.
Insets show the fluctuations of HFACF and thermal conductivities with dashed lines indicating average values estimated from 1000 step.}
\end{figure}
We first presented the thermal conductivity of perfect jennite calculated by GK approach using the $3\times 4\times 3$ supercell model.
Figure~\ref{fig_green_kubo} shows the HFACF along the [100], [010] and [001] directions, and thus the principal elements of thermal conductivity tensor at temperature of 300 K.
After approximately 200 fs of $NVE$ simulation at equilibrium state, the HFACF was found to fluctuate around zero while the thermal conductivity to converge into a certain value.
The principal values were determined to be $K_1=1.125$, $K_2=1.393$ and $K_3=0.921$ W/m$\cdot$K along the [100], [010] and [001] directions respectively, and the average volumetric value was estimated to be $K_\text{v}=1.146$ W/m$\cdot$K.
As expected earlier, the thermal conductivity of jennite was confirmed to be anisotropic like $K_2 > K_1 > K_3$, which is different from $K_1 > K_2 > K_3$ for C$-$S$-$H predicted by MD~\cite{Qomi1}, possibly due to differences in model and force field.
The reason of such anisotropy is that fairly heat conductive Si$-$O bonds attribute to higher thermal conduction along the [010] direction (the so-called Si$-$O dreierketten chains run along the [010] direction~\cite{Bonaccorsi-jen}), whereas in the [001] direction water molecules and looser Ca$-$O bonds scatter phonons, resulting in lower thermal conduction in this direction~\cite{Dongshuai, Qomi1}.

\subsection{\label{sub_length}Influence of system length on thermal conductivity}
In the previous works, it has been found that in the MP method the accuracy of thermal conductivity calculation severely depends on the system length, whereas the effect of its cross-section area is independent for the case of amorphous nanoporous silica.~\cite{Coquil}.
Therefore, we tested the influence of simulation box length on the thermal conductivity of bulk jennite to determine the proper length.
In this test, the perfect jennite model was used.
Due to the anisotropy of thermal conductivity, the jennite model was repeated from two to ten times in the [100], [010] and [001] directions, yielding the simulation boxes with a range of system length from $\sim$60 to $\sim$300 \AA, cross-section area of about $30\times 30$ \AA$^2$, and number of atoms from 4968 to 24 840.

\begin{figure}[!t]
\centering
\includegraphics[clip=true,scale=0.1]{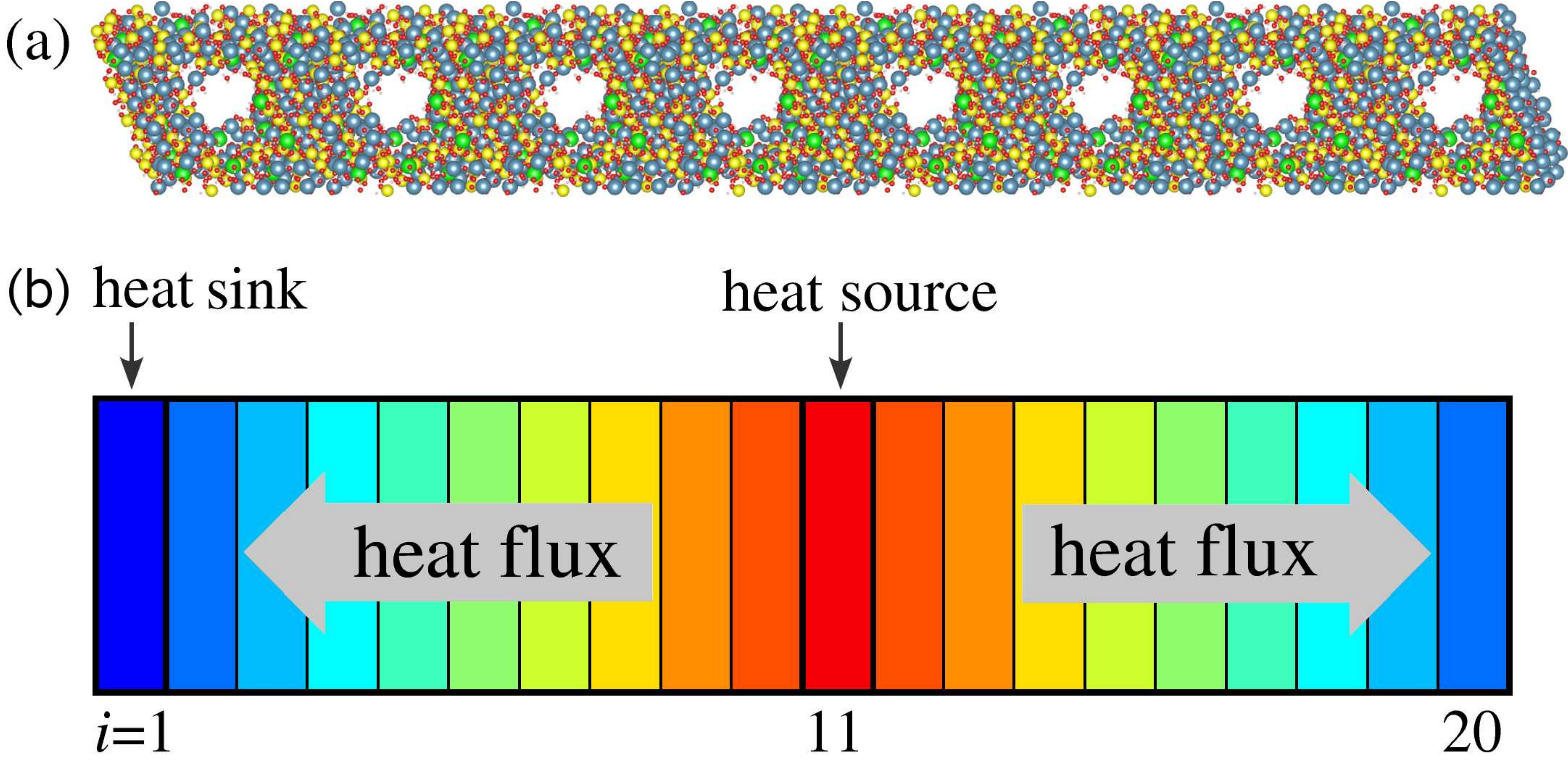}
\includegraphics[clip=true,scale=0.49]{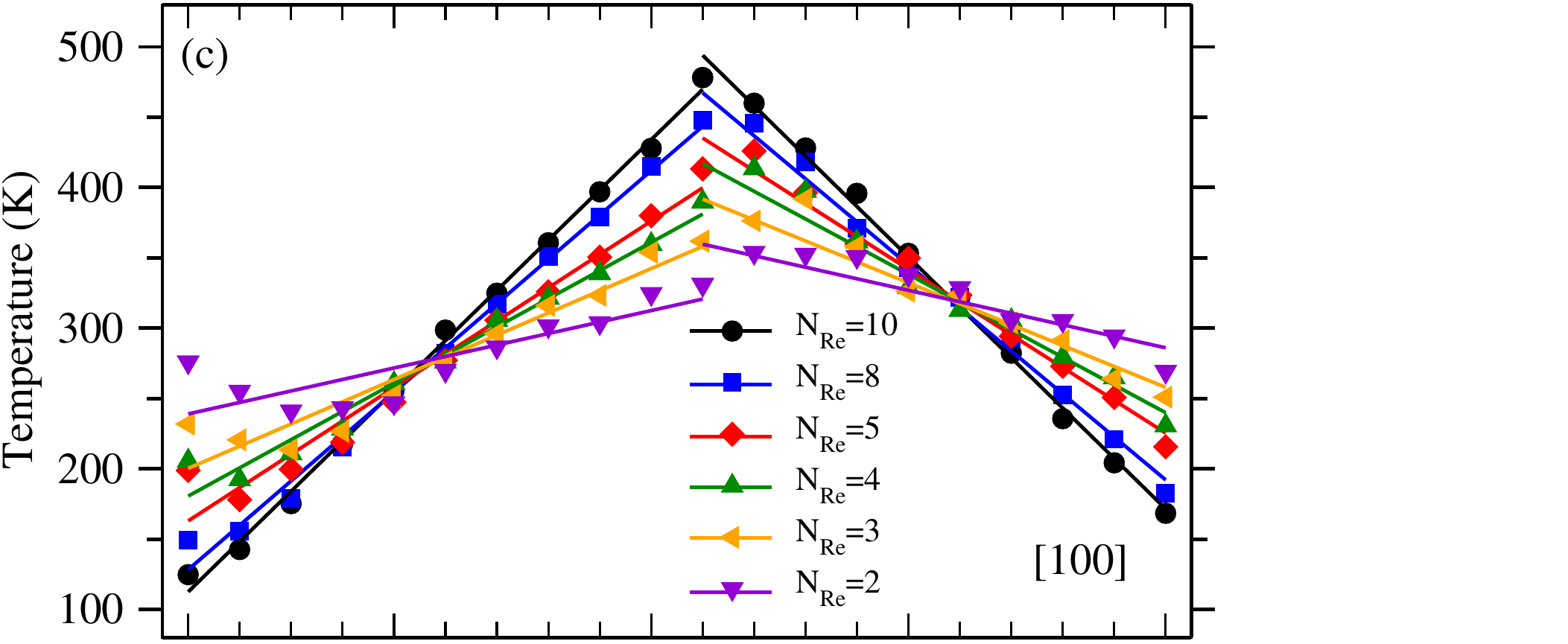}
\includegraphics[clip=true,scale=0.49]{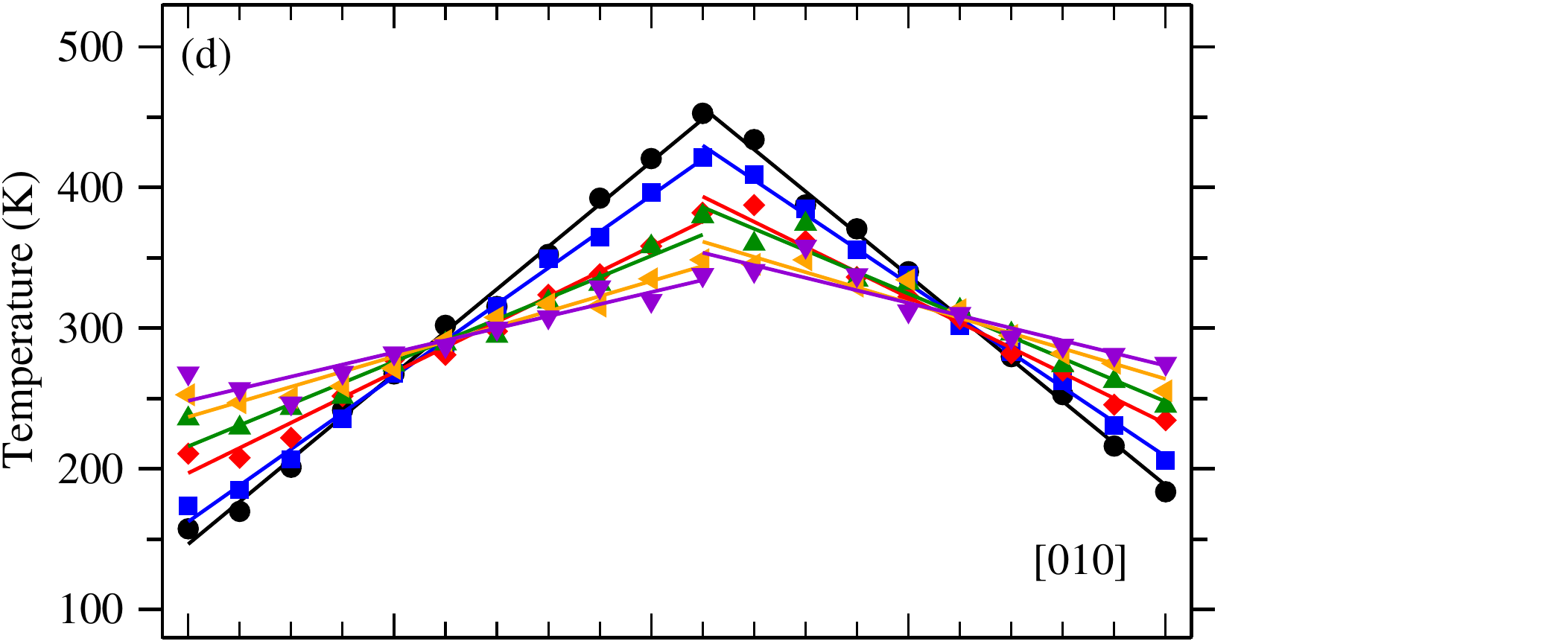}
\includegraphics[clip=true,scale=0.49]{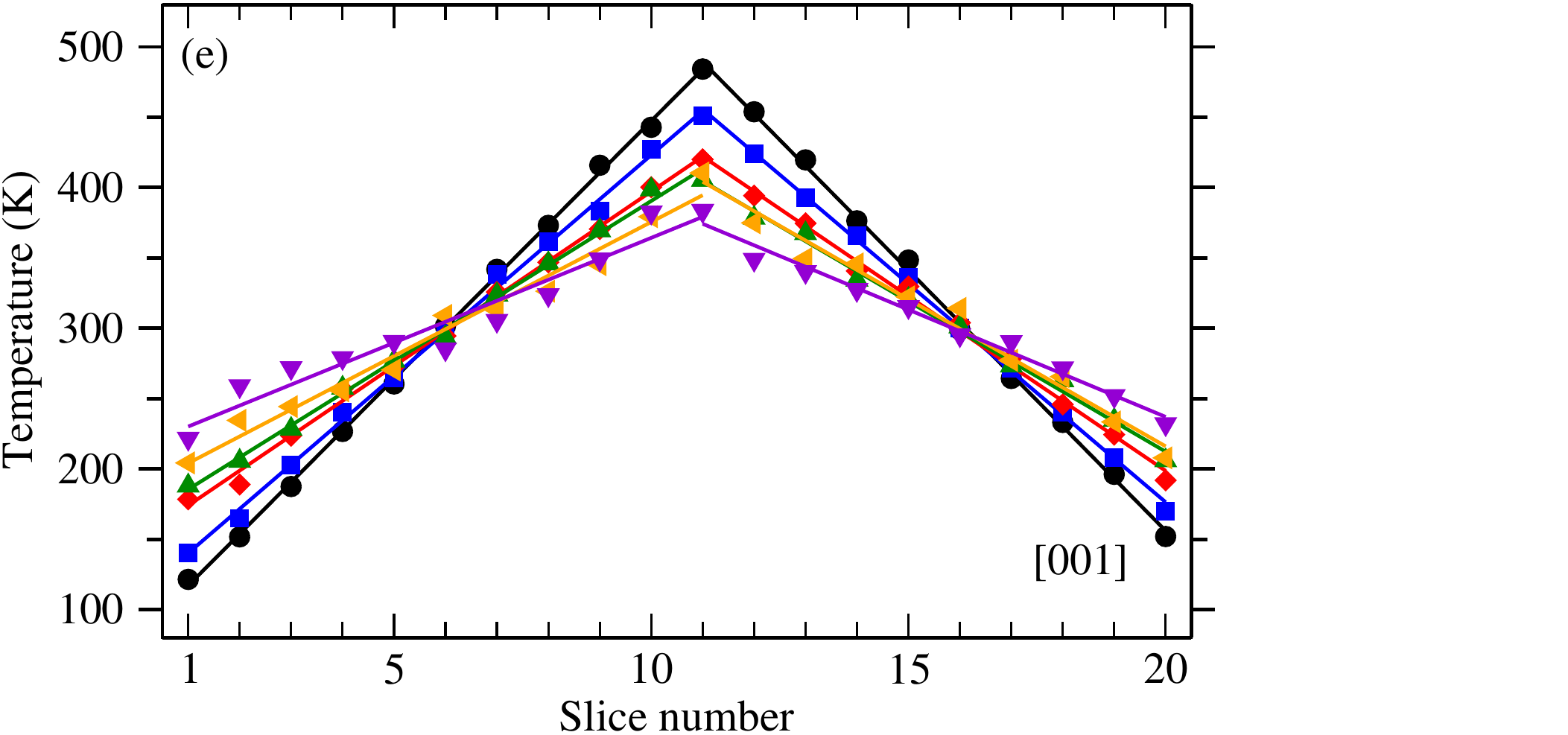}	
\caption{\label{fig3}(a) The simulation box created by repeating the supercell (porosity 24.4\%) eight times in [100] direction. (b) Sketch of the simulation box divided into 20 slices, of which the first and 11th slices are for heat sink and source. Temperature profiles as increasing the system length (number of repeated jennite cells $N_{\text{Re}}$) in (c) [100], (d) [010], and (e) [001] directions at 300 K.}
\end{figure}
As depicted in Fig.~\ref{fig3}(b), the simulation boxes were divided into 20 plane slices perpendicular to the repeating direction (heat flux propagation direction), of which the 11th slice plays the role of heat source and the first slice is for heat sink.
It is worth noting that this configuration is due to the periodic boundary condition in all the three directions, while under the fixed boundary condition the first and last slices can be used for heat source and sink~\cite{Zhu}.
Temperature was fixed at 300 K through $NVT$ simulation, and subsequently $NVE$ runs were performed with a time step of 0.1 fs during a total time of 200 ps.
Figures~\ref{fig3}(c)-(e) show temperature profiles in the simulation boxes with increasing lengths along the heat flux directions of [100], [010] and [001].
It was observed that, when the system length was not enough long, the temperature reduction going from the heat source to the heat sink slices did not follow the straight line but rather meandering path, raising difficulty and ambiguity in measuring the temperature gradient.
For the system lengths over $\sim$150 \AA~(extending supercell model five times), the temperature profile became almost the straight line.

\begin{figure}[!t]
\centering
\includegraphics[clip=true,scale=0.5]{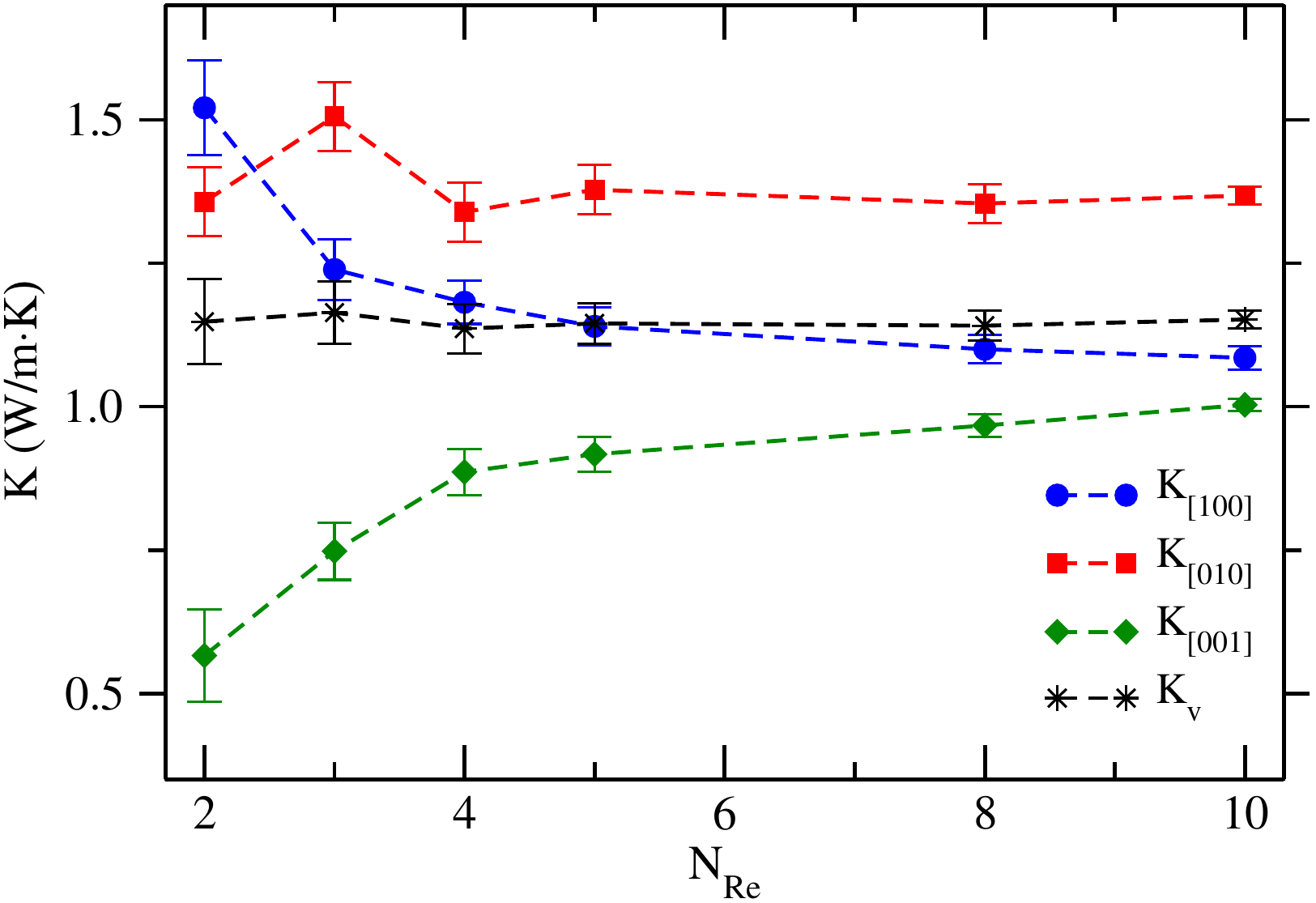}
\caption{\label{fig4}Thermal conductivity of perfect jennite as increasing the system length, obtained by applying the non-equilibrium M\"{u}ller-Plathe method.}
\end{figure}

\begin{table}[!b]
\small
\caption{\label{tbl_size}Principal ($K_1$, $K_2$, $K_3$) and volumetric averaged ($K_\text{v}$) thermal conductivities of jennite crystal as increasing the system length, i.e. number of repeating cells $N_{\text{Re}}$.}
\begin{tabular}{ccccc}
\hline
$N_{\text{Re}}$ & $K_1$ (W/m$\cdot$K) & $K_2$ (W/m$\cdot$K) & $K_3$ (W/m$\cdot$K) & $K_\text{v}$ (W/m$\cdot$K) \\
\hline
2   & 1.521  & 1.357  & 0.566   & 1.148  \\
3   & 1.239  & 1.506  & 0.748   & 1.164  \\
4   & 1.182  & 1.339  & 0.886   & 1.136  \\
5   & 1.140  & 1.378  & 0.917   & 1.145  \\
8   & 1.100  & 1.354  & 0.967   & 1.141  \\
10  & 1.085  & 1.368  & 1.003   & 1.152  \\
\hline
\end{tabular} 
\end{table}
In Table~\ref{tbl_size}, we summarize the principal and volumetric averaged thermal conductivities of jennite crystal as increasing the system length.
As can be seen in Fig.~\ref{fig4}, the principal thermal conductivities converged into the reliable values over the system length of $\sim$150 \AA, which is comparable with the previous value of 100 \AA~for amorphous silica~\cite{Coquil}.
For the sake of high reliability, we selected the appropriate system length as about 240 \AA~(repeating jennite model 8 times, 19 872 atoms) by considering the computational cost and statistical error.

When compared with our previous result by GK approach, the obtained thermal conductivities, $K_1=1.100\pm0.025$ W/m$\cdot$K, $K_2=1.354\pm 0.034$ W/m$\cdot$K, $K_3=0.967\pm 0.020$ W/m$\cdot$K and $K_\text{v}=1.141\pm 0.026$ W/m$\cdot$K, are in good agreement each other.
Due to a lack of experimental data for jennite, we compared our results with the previous MD works for similar materials.
For C$-$S$-$H, Qomi {\it et al}.~\cite{Qomi1} obtained $0.98\pm 0.2$ W/m$\cdot$K, which is comparable with ours.
Wu {\it et al}.~\cite{Wu} also reported the similar value of concrete hydration product as 1.015 W/m$\cdot$K through the volumetric weighted average of the alite and belite.
These indicate that our models, adopted force field, and computational method are sufficiently accurate for study of porous jennite in the following subsections.

\subsection{\label{sub_poro}Porosity dependence of thermal conductivity}
We further investigated the porous jennite with porosities from 0\% to 72.22\% at fixed temperature of 300 K, comparing with the different models for describing the effective thermal conductivity of porous material and with the experimental data.
It should be noted that introducing pores into the perfect crystal or media reduces the thermal conductivity substantially, and thus, porous media acts as a thermal insulator or a barrier to heat removal.
As mentioned above, the simulation boxes extended by 8 times in each direction were employed (see Fig.~\ref{fig3}(a) for porous jennite with a porosity of 24.39\%).
However, for the case of 72.22\% porosity, the number of atoms was too small (5 520 atoms) and thus the width and height of simulation box were again doubled, keeping the length unchanged, so that the number of atoms became enough large (22 080).

Before presenting the results of this work, let us consider the previously suggested models that relate the thermal conductivity of porous media to the porosity.
The Garnett model was first derived to determine the electrical conductivity of a binary composite, in which spherical inclusions with random size were dispersed, as follows~\cite{Maxwell},
\begin{equation}
\label{eq_maxwell}
K^\text{G}=K_b\frac{2K_b+K_p-2p(K_b-K_p)}{2K_b+K_p+p(K_b-K_p)}
\end{equation}
where $K_b$ and $K_p$ are the thermal conductivities of bulk and pore, and $p$ is the porosity. 
Landauer~\cite{Landauer} proposed the coherent potential (CP) model to estimate the conductivity of a binary composite made of matrix and spherical inclusions as follows,
\begin{multline}
\label{eq_cp}
K^{\text{CP}}=\frac{1}{4} \bigg \{ (3p-1)K_p+(2-3p)K_b + \\ 
\left[\Big((3p-1)K_p+(2-3p)K_b\Big)^2+8K_bK_p \right]^\frac{1}{2} \bigg \}
\end{multline}
By using the CP model, Cahill and Allen~\cite{Cahill} well predicted the thermal conductivity of porous material with a porosity of 30\%.
The Russel model was developed to derive the effective thermal conductivity of a dry porous material, in which cubical inclusions with uniform pores were evenly distributed~\cite{Russell},
\begin{equation}
\label{eq_russell}
K^\text{R}=K_b\frac{vp^{2/3}+1-p^{2/3}}{v(p^{2/3}-p)+1+p-p^{2/3}} 
\end{equation}
where $v=K_p/K_b$.
The Russell model assumed a parallel heat flow, whereas the Frey model treated a series heat flow~\cite{Frey},
\begin{equation}
\label{eq_frey}
K^\text{F}=K_b\frac{v(1-p^{1/3}+p)+p^{1/3}-p}{v(1-p^{1/3})+p^{1/3}} 
\end{equation}
The Parallel and Series models were developed to give the maximum and minimum of thermal conductivity of a porous material because planes of constituent phases are in parallel and in series with the heat flow direction, as follows,
\begin{equation}
\label{eq_parallel}
K^\text{P}=K_b(1-p)+K_pp
\end{equation}
\begin{equation}
\label{eq_series}
K^\text{S}=\frac{1}{(1-p)/K_b+p/K_p}
\end{equation}
Meanwhile, Qomi {\it et al}.~\cite{Qomi1} derived the self-consistent (SC) model to estimate the homogenized thermal conductivity of matrix-inclusion system, as follows,
%
%
\begin{equation}
\label{eq_sc}
K_{v}^\text{SC}=\frac{\sum_{s=1}^{n_p}{f_sK_v^sB_s^\text{sph}}}{\sum_{s=1}^{n_p}{f_sB_s^\text{sph}}}
\end{equation}
where $K_v^s$ is the volumetric thermal conductivity of spherical inclusion, $f_s$ is the volume fraction between the matrix and inclusion, and $B_s^\text{sph}=3K_v^\text{SC}/(2K_v^\text{SC}+K_v^s)$ is the spherical localization factor of the $s$th phase.

\begin{table}[!b]
\small
\caption{\label{tbl_kp}Principal ($K_1$, $K_2$, $K_3$) and volumetric averaged ($K_\text{v}$) thermal conductivities of the porous jennite with various porosities ($P_\rho$) at 300 K.}
\begin{tabular}{l@{\hspace{6pt}}c@{\hspace{6pt}}c@{\hspace{6pt}}c@{\hspace{6pt}}c}
\hline
$P_\rho$ (\%) & $K_1$ (W/m$\cdot$K) & $K_2$ (W/m$\cdot$K) & $K_3$ (W/m$\cdot$K) & $K_{\text{v}}$ (W/m$\cdot$K)\\
\hline
0.00  & 1.100 & 1.354  & 0.967 & 1.141  \\
4.79  & 0.996 & 1.253  & 0.867 & 1.039  \\
10.39 & 0.932 & 1.159  & 0.806 & 0.966  \\
15.99 & 0.828 & 0.986  & 0.732 & 0.849  \\
24.39 & 0.660 & 0.770  & 0.671 & 0.701  \\
32.79 & 0.525 & 0.684  & 0.545 & 0.585  \\
52.39 & 0.200 & 0.384  & 0.338 & 0.307  \\
72.22 & 0.013 & 0.273  & 0.146 & 0.144  \\
\hline
\end{tabular} 
\end{table}
\begin{figure}[!t]
\centering
\includegraphics[clip=true,scale=0.5]{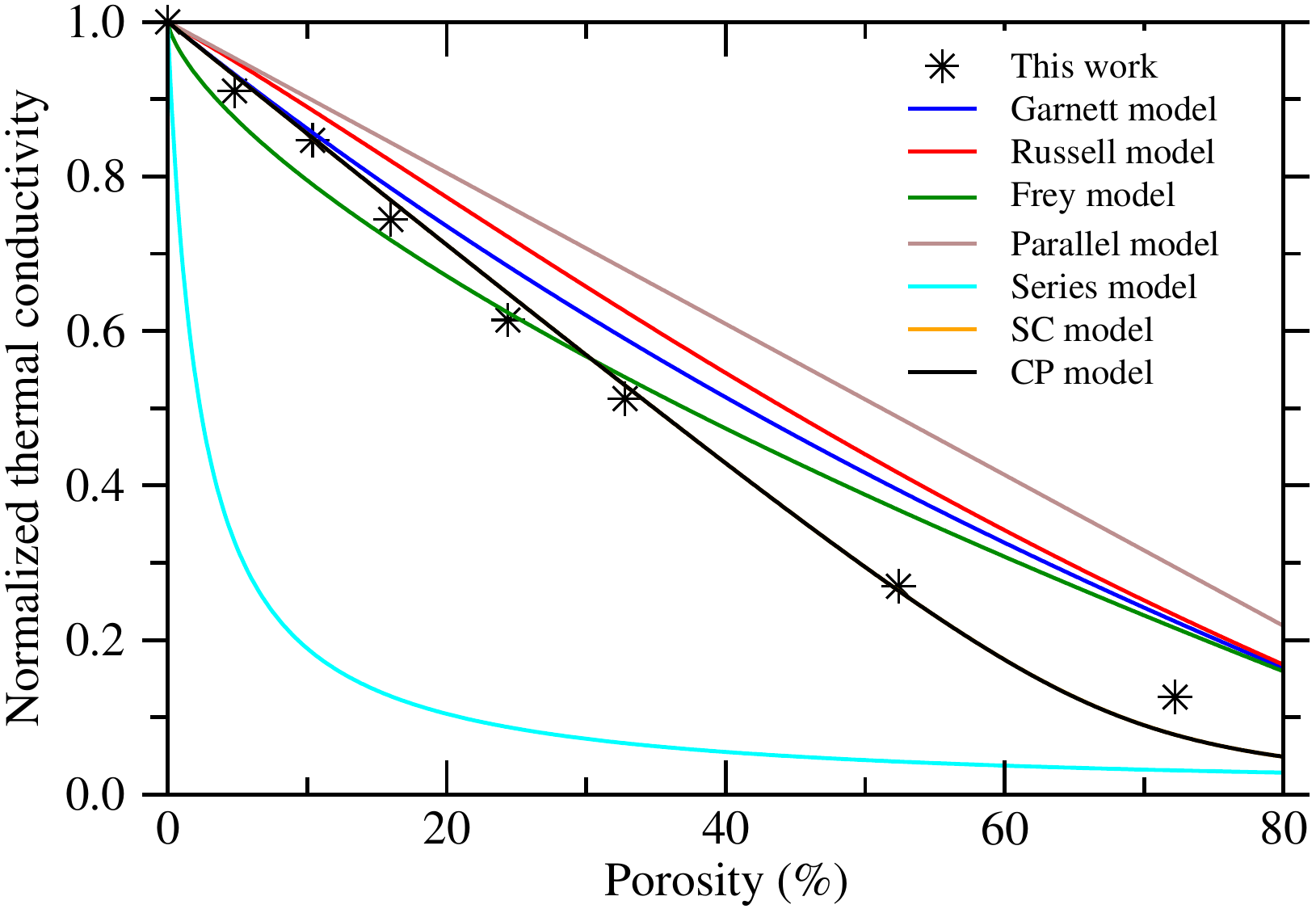}
\caption{\label{fig_kp}Normalized thermal conductivity of porous jennite as a function of porosity obtained from the suggested models and MD simulation in this work.}
\end{figure}

In Table~\ref{tbl_kp}, we present the thermal conductivities of the porous jennite with different porosities from 0 to 72.22\% calculated by applying the MP method at 300 K.
Figure~\ref{fig_kp} shows normalized thermal conductivity of the porous jennite as a function of porosity, comparing with the effective thermal conductivities of porous media derived from the various models described above.
As expected, the thermal conductivity of porous jennite was observed to decrease as increasing the porosity, since the thermal conductivity of air (0.026 W/m$\cdot$K~\cite{Lide}) is much smaller than that of jennite.
As can be seen in Fig.~\ref{fig_kp}, although all models exhibit decreasing tendency of effective thermal conductivity as increasing the porosity, the CP and SC models (they are almost identical each other) were found to be in quantitative agreement with our MD simulation results.
It should be noted that an iterative procedure is required in the calculation with the SC model, and thus, in order to calculate the thermal conductivity of porous jennite by using the SC model we made use of the result from the Parallel model as the initial value and iterated the calculation for 10 times during the self-consistent cycle, reaching convergence after 7 or 8 times.
For the case of amorphous \ce{SiO2}, the CP model has also been found to reasonably produce its thermal conductivity for a range of porosities such as 10 $\sim$ 35\%~\cite{Coquil} and 0.54 $\sim$ 24.15\%~\cite{Zhu}.
Furthermore, Santos~\cite{Santos} revealed that the thermal conductivity of conventional aluminous refractory concrete exhibited the decreasing tendency as increasing the porosity from 0 to $\sim$36.37\% at various temperatures, nearly according to that from the CP model.
Meanwhile, Coquil {\it et al}.~\cite{Coquil} found a slight overestimation of the CP model compared with the MD result, whereas Zhu {\it et al}.~\cite{Zhu} found an underestimation when the porosity was 24.15\%.
But in our study the MD results were in good agreement with those from the CP model over the porosity range from 0 to 52.39\% and not ruled out for the porosity of 72.22\%.
Based on this analysis, it can be concluded that the CP model is able to well predict the thermal conductivity of not only amorphous porous silica but also porous cement based materials.

\begin{figure}[!t]
\centering
\includegraphics[clip=true,scale=0.5]{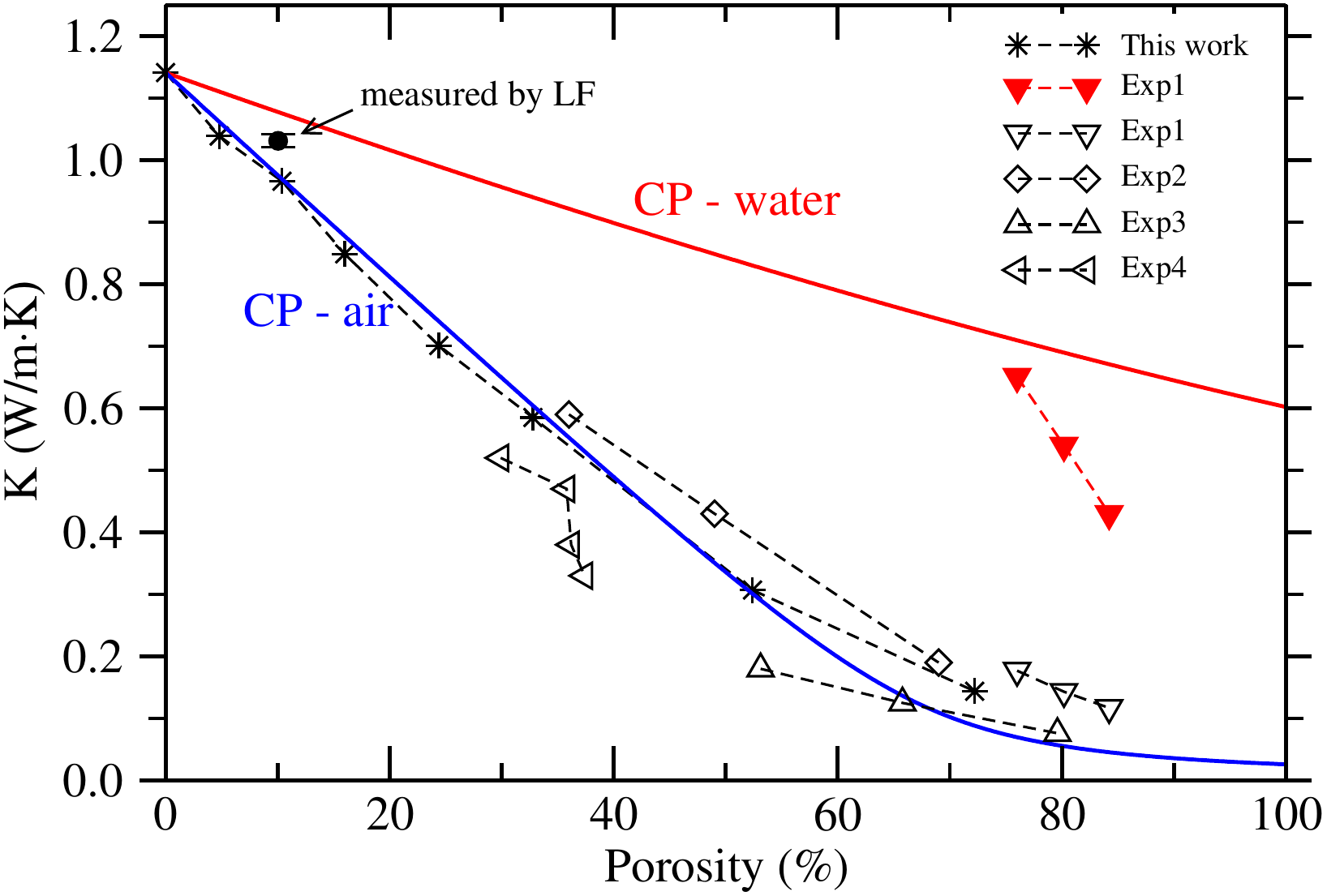}
\caption{\label{fig6}Thermal conductivity of cement hydration products including jennite with their pores filled with air or water, comparing with experimental data.
Exp1, 2, 3, 4 are from Ref.~\cite{QingJin, Awang, Batool, Gencel}.
They are well placed between the blue line for air filling and the red line for water filling estimated by using the CP model.}
\end{figure}
Next, we made a careful analysis of our MD result in comparison with those from the CP model, in which the pores were assumed to be filled with air or water, and from experiment.
Whether the pores of porous cement hydration product would be filled with air or water (or even vacuum) depends on its hardening process condition.
If the pores are assumed to be filled with water, the porous jennite has the maximal thermal conductivity, whereas filling with air will give the minimal values, as obtained by the CP model shown in Fig.~\ref{fig6}.
It is worth noting that for the case of water filling the thermal conductivity of porous jennite should decrease with increasing the porosity as well, due to smaller conductivity of water (0.6062 W/m$\cdot$K~\cite{Lide}) than jennite.
According to the work of Constantinides and Ulm~\cite{Constantinides}, the porosities of high and low density C$-$S$-$H were found to be 24 and 36\% respectively, and therefore, if hydrated cement paste would be composed of only jennite, its thermal conductivity should be ranged from 0.553 W/m$\cdot$K with air filling to 0.992 W/m$\cdot$K with water filling on the CP model lines.

\begin{table}[!b]
\small
\caption{\label{tbl_kwc}Thermal conductivity ($K$) of the cement hydration product with different water to cement ratios (w/c)}
\begin{tabular}{lll}
\hline
                        & $K$ (W/m$\cdot$K)  & w/c\\
\hline
This work               & 1.030              & --  \\
Yoon {\it et al}.~\cite{Yoon} & 1.28               & 0.4  \\
Xu {\it et al}.~\cite{Xu}     & 0.53               & 0.35  \\
Bentz~\cite{Bentz}      & 0.90 $\sim$ 1.05   & 0.3, 0.4  \\
Hansen {\it et al}.~\cite{Hansen} & 0.88 (early)   & 0.5  \\
		                & 0.78 (late)        &  \\
Mounanga {\it et al}.~\cite{Mounanga} & 1.0 (fresh) & 0.348  \\
	                    & 1.07 (28 days)     &   \\
Wu {\it et al}.~\cite{Wu}     & $0.9568\pm 0.0257$ & 0.45  \\
\hline
\end{tabular} \\
\footnotesize
In this work, laser flash method was used to measure thermal conductivity. \\
Data of Wu {\it et al}.~\cite{Wu} was from calculation, and others were from experiment.
\end{table}
In our previous work~\cite{yucj19mcp}, we demonstrated that when applying the multi-functional composite coating material~\cite{Ri18wipo} made from silica/alumina-rich minerals to the concrete wall, high dense ceramic layer with a thickness of $\sim$1 mm and a porosity of 10.05\% can be formed on the wall surface by two-step hydration of cement and mineral powders.
By using the laser flash method, we measured the thermal conductivity of this ceramic layer to be 1.030 W/m$\cdot$K at 223 K (see Table S2), which is placed between the red and blue lines as shown in Fig.~\ref{fig6}.
In Table~\ref{tbl_kwc}, we summarize the thermal conductivity of cement hydration products with different water to cement ratios (w/c), measured from experiment or calculated from theory.
Wu {\it et al}.~\cite{Wu} calculated the thermal conductivity of hardened cement paste to be $0.96\pm 0.03$ W/m$\cdot$K when its hydration degree was 0.95 and w/c =0.45.
When compared with the available experimental data of cement paste with different w/c ratios such as $0.90\sim 1.05$ W/m$\cdot$K with the hydration degree of about $0.3\sim 0.8$, 1.28 W/m$\cdot$K~\cite{Yoon}, 0.78 and 0.88 W/m$\cdot$K for early and late stage of hydration process~\cite{Hansen}, 0.53 W/m$\cdot$K~\cite{Xu}, and 1.0 and 1.07 W/m$\cdot$K for fresh and after 28 days~\cite{Mounanga}, our predicted values with MD simulation can be said to be well suited with them.


\subsection{\label{sub_temp}Temperature dependence of thermal conductivity}
At the final stage, we investigated the thermal conductivity of porous jennite as increasing temperature from 240 K to 560 K with a step of 60 K for details and from 300 K to 1300 K with a step of 200 K.
For the sake of simplicity without losing generality, we selected the three different typical porosities of 0\%, 15.99\% and 32.79\%.
We followed the above-mentioned procedure to calibrate the thermal conductivity: doing $NPT$ simulation of perfect $3\times4\times3$ jennite model for 100 ps, creating the pore with a radius corresponding to the porosity and running $NVT$ for 400 ps, constructing simulation box by repeating 8 times the model and running $NVT$ for 100 ps, and finally performing $NVE$ run for 200 ps with imposing the heat flux, all these with a time step of 0.1 fs.
These processes were repeated at different temperatures.

\begin{table}[!b]
\small
\caption{\label{tbl_tempcell}The unit cell parameters and density of jennite crystal at different temperatures, obtained by MD simulation.}
\begin{tabular}{l@{\hspace{6pt}}c@{\hspace{6pt}}c@{\hspace{6pt}}c@{\hspace{6pt}}c@{\hspace{6pt}}c@{\hspace{6pt}}c@{\hspace{6pt}}}
\hline
                    & 300 K  & 500 K  & 700 K  & 900 K  & 1100 K & 1300 K\\
\hline
$a$ (\AA)           & 10.865 & 11.600 & 12.691 & 13.666 & 12.647 & 14.238 \\
$b$ (\AA)           & 7.402  & 7.627  & 7.188  & 7.246  & 6.183  & 7.312 \\
$c$ (\AA)           & 10.946 & 10.474 & 11.014 & 10.832 & 15.031 & 25.811 \\
$\alpha$ ($^\circ$) & 102.06 & 103.66 & 110.64 & 98.35  & 89.83  & 90.78 \\
$\beta$ ($^\circ$)  & 94.57  & 93.95  & 84.53  & 93.47  & 99.12  & 89.64 \\
$\gamma$ ($^\circ$) & 110.33 & 111.55 & 110.63 & 114.55 & 104.31 & 106.85 \\
V (\AA$^3$)         & 796.1  & 824.9  & 879.4  & 956.6  & 1123.6 & 2571.8 \\
$\rho$ (g/cm$^{3}$) & 2.218  & 2.141  & 2.008  & 1.846  & 1.633  & 0.687 \\
\hline
\end{tabular} 
\end{table}
Table~\ref{tbl_tempcell} lists the unit cell parameters and density of perfect jennite crystal obtained by doing $NPT$ simulations as increasing temperature from 300 K to 1300 K.
It was observed that as increasing temperature the unit cell volume increases monotonically while the mass density decreases, although the lattice constants varies irregularly.
It should be noted that at 1300 K the density was too small to think of crystal; it was actually transformed to liquid state.
Moreover, as revealed by Yu and Kirkpatrick~\cite{Yu}, the jennite crystal is known to be transformed to wollastonite and larnite above 1073 K.
That is why we did not perform further simulation at 1300 K.
Another point is that since the volume of jennite model increases with increasing temperature, the radius of spherical region for creating the pore should be adjusted to meet the target porosity.

\begin{figure}[!t]
\centering
\includegraphics[clip=true,scale=0.5]{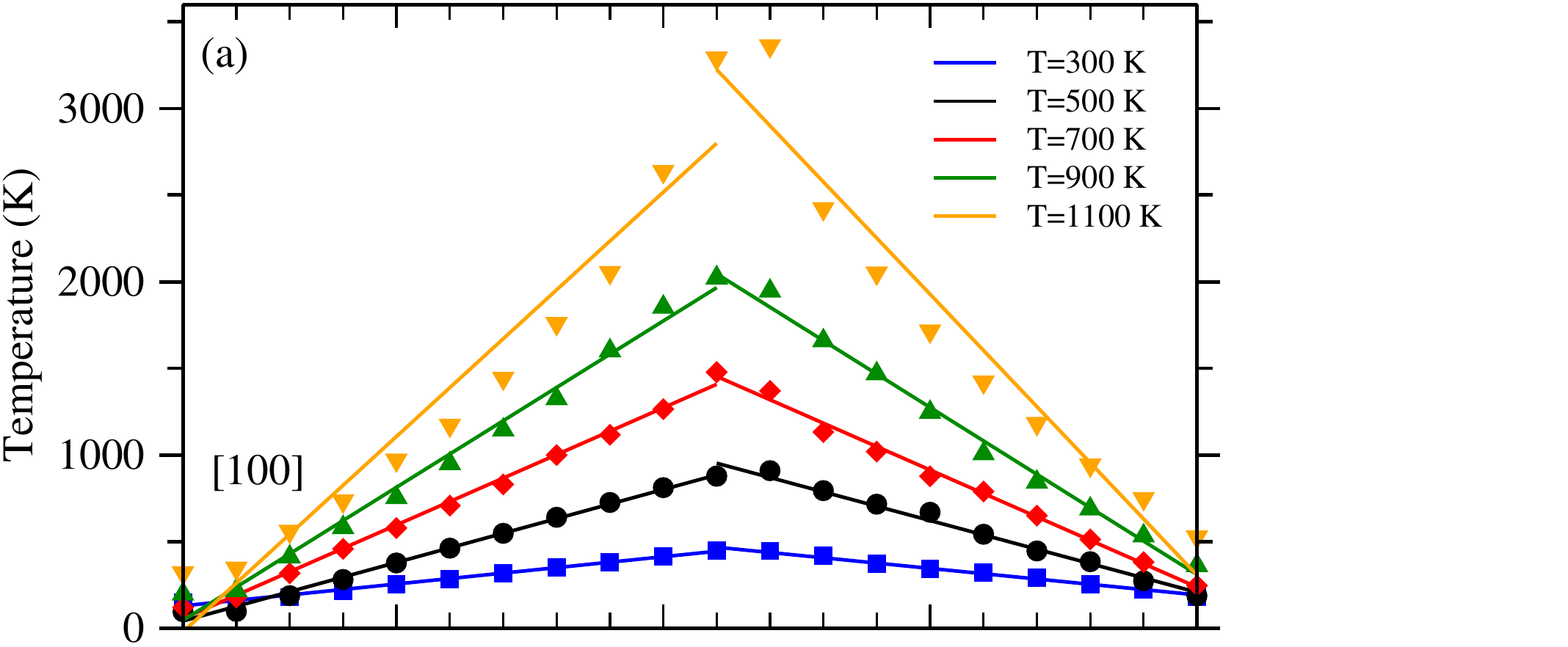}
\includegraphics[clip=true,scale=0.5]{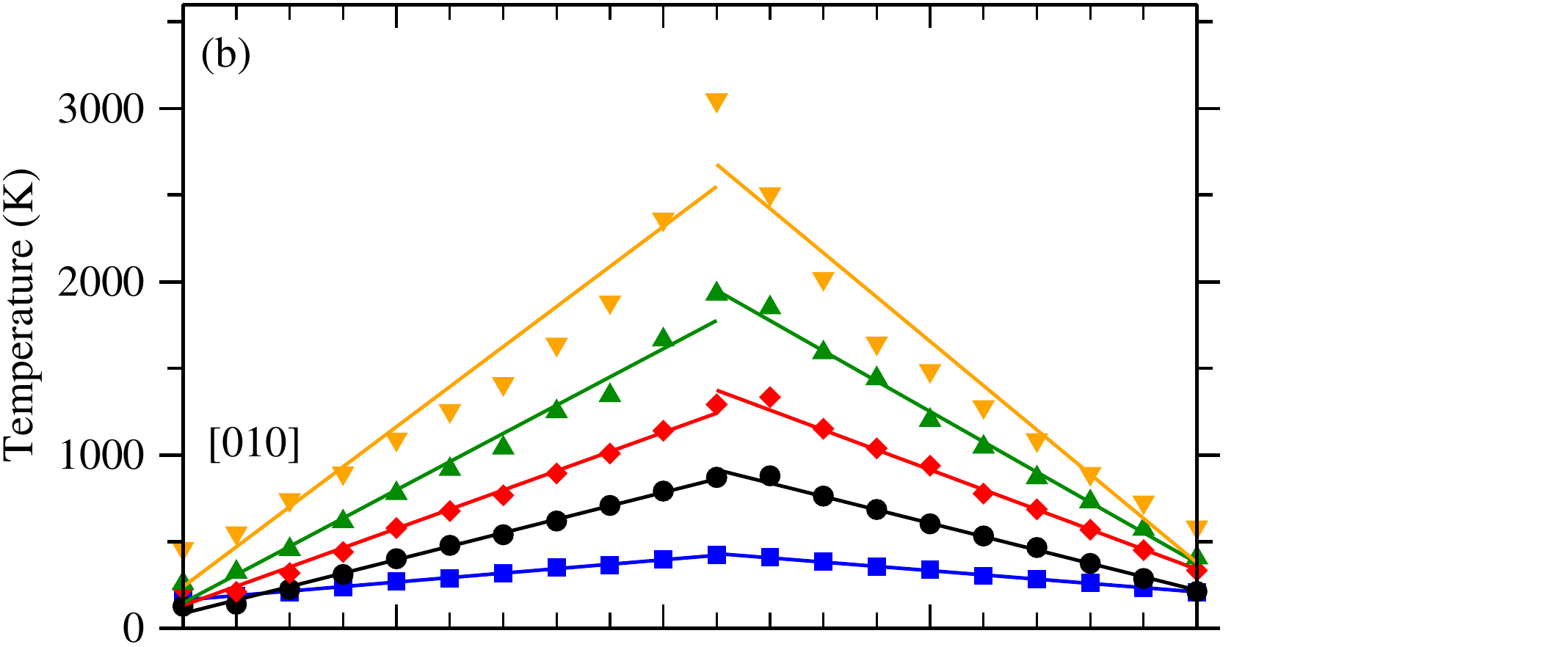}
\includegraphics[clip=true,scale=0.5]{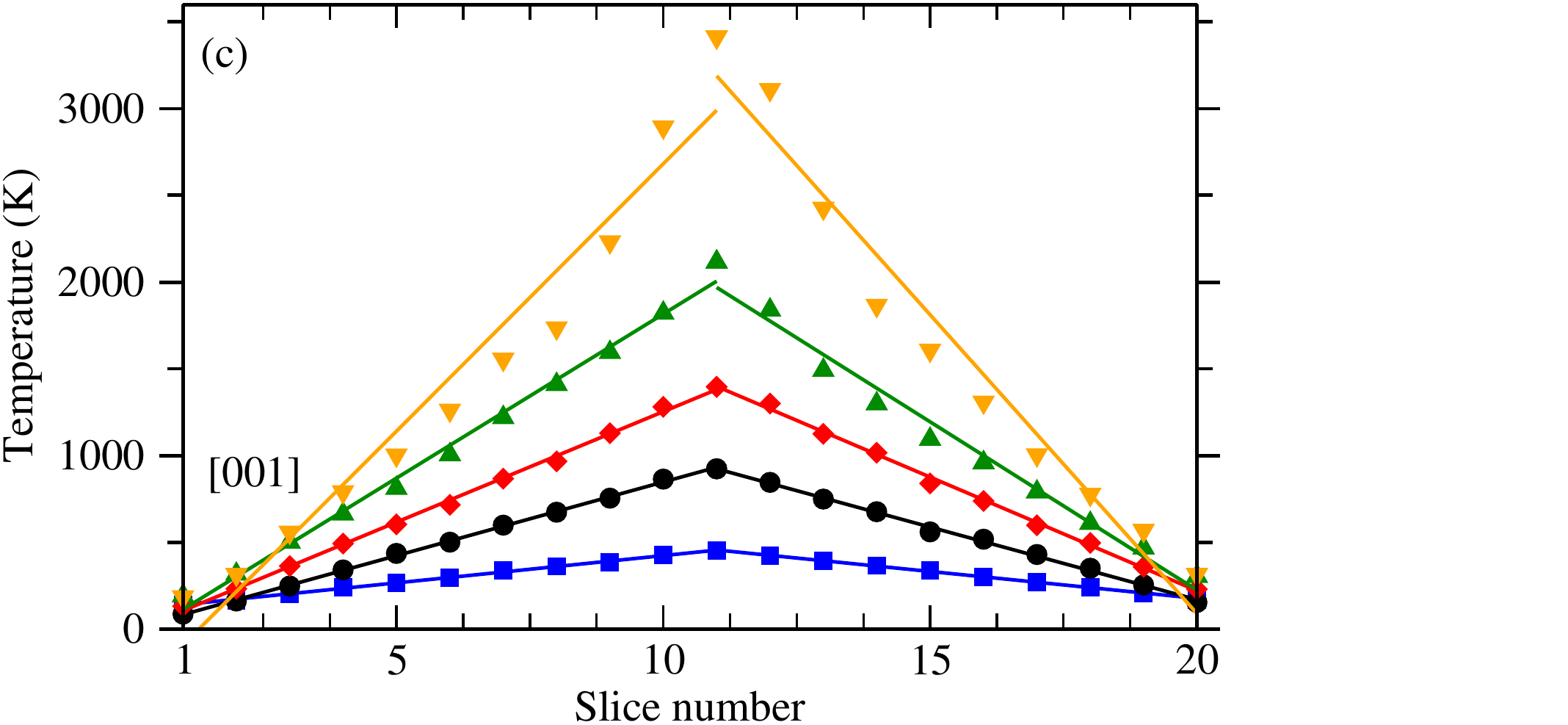}
\caption{\label{fig7}Temperature profiles in simulation boxes of perfect jennite crystal along (a) [100], (b) [010] and (c) [001] directions as increasing temperature from 300 K to 1100 K with a step of 200 K.}
\end{figure}
Figure~\ref{fig7} shows the temperature profiles in the simulation boxes of perfect jennite along the principal crystallographic directions.
At moderate temperatures, one can see a good linearity of temperature change along the slices, which makes it easy to measure the temperature gradients, while an inappreciable dispersion of temperatures was found at 1100 K.

\begin{figure}[!t]
\centering
\includegraphics[clip=true,scale=0.5]{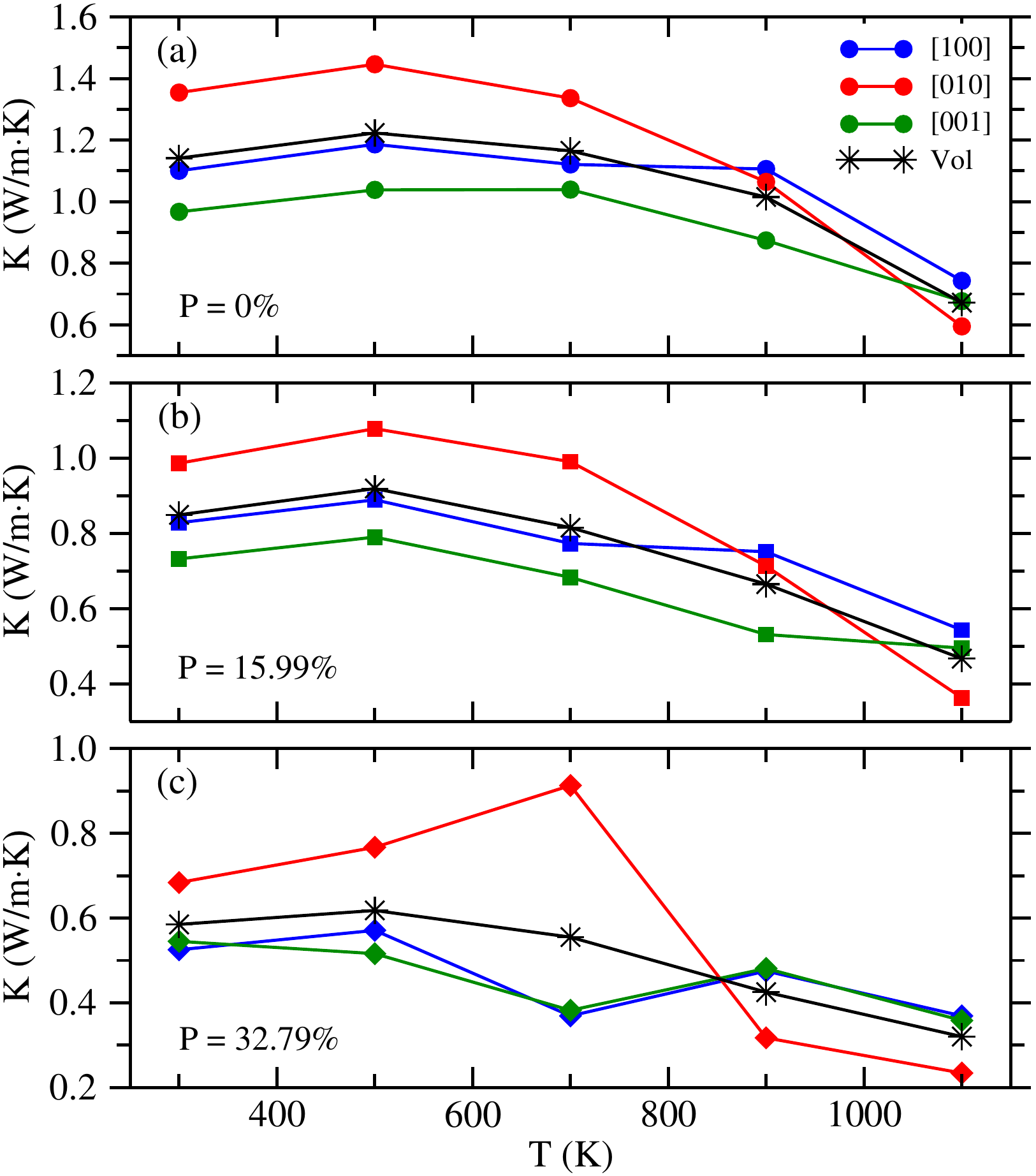}
\includegraphics[clip=true,scale=0.5]{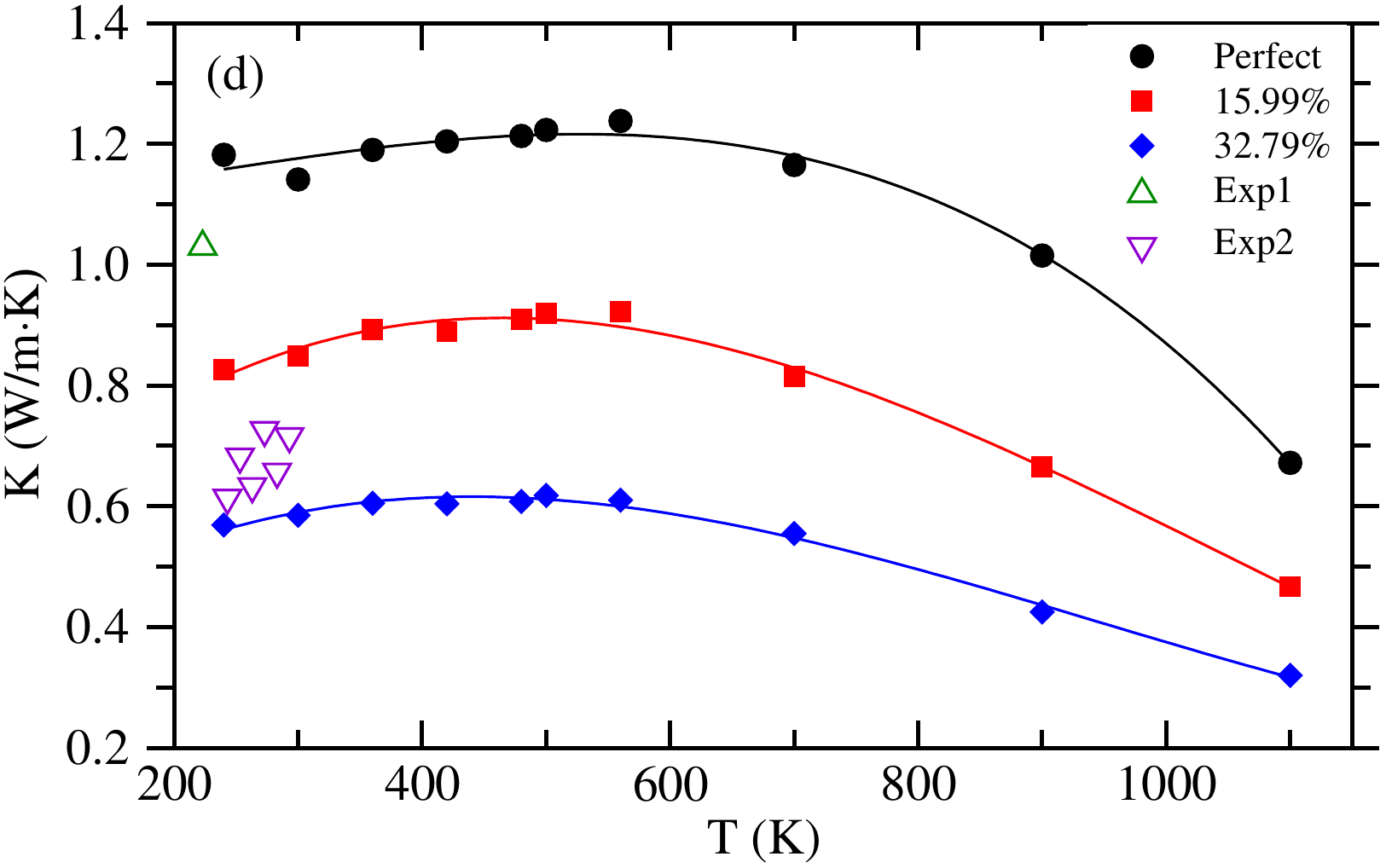}
\caption{\label{fig8}Thermal conductivities of porous jennite with various porosities of (a) 0\%, (b) 15.99\%, and (c) 32.79\% along the [100], [010] and [001] directions as increasing temperature from 300 K to 1100 K with a step of 200 K, where Vol means the volumetric averaged value.
(d) Volumetric thermal conductivities as functions of temperature, merged those from 240 K to 560 K with a step of 60 K and those from 300 K to 1100 K with a step of 200 K, in comparison with our measurement (Exp1, 2) performed by using the laser flash method.
Solid lines indicate the 3rd-order polynomial interpolation.}
\end{figure}
In Fig.~\ref{fig8}, we show the temperature dependence of thermal conductivities of porous jennite with porosities of 0, 15.99 and 32.79\%.
For the three cases of porosities, the principal thermal conductivities, especially volumetric averaged conductivities, increase to 500 K and decrease after that.
Also, it was found that the [100] element of thermal conductivity became the biggest after 900 K in all the three cases.
Figure~\ref{fig8}(d) shows the volumetric thermal conductivities calculated in the detailed temperature range from 240 K to 560 K as well as in the temperature range from 300 K to 1100 K in comparison with our experimental values measured for the ceramic coating layer on the concrete wall by using the laser flash method.
We note that our multifunctional composite building material was developed for use of thermal insulation, moisture barrier, sterilization of buildings~\cite{Ri18wipo}.
Our experimental values were found to mostly locate between the porosity of 15.99\% and 32.79\%.
Since the porosity of this ceramic coating layer was measured to be about 10\%~\cite{yucj19mcp}, these are slightly deviated from the predicted values, possibly due to not a pure jennite.
Based on our calculation result, we suggested the 3rd-order polynomial relation between the thermal conductivity of porous jennite and temperature by
\begin{equation}
\label{eq_intp}
K(T)=c_0+c_1T+c_2T^2+c_3T^3
\end{equation}
where $c_i$ is the coefficient. Table~\ref{tbl_intp} shows them resulting from interpolation process and in Fig.~\ref{fig8}(d) we see that the 3rd-order interpolation fits well with our calculation data.
\begin{table}[!th]
\small
\caption{\label{tbl_intp}Coefficients of the 3rd-order polynomial relation between the thermal conductivity of porous jennite and temperature.}
\begin{tabular}{cccrr}
\hline
$P$ (\%) & $c_0$  & $c_1(\times10^{-3})$  & $c_2(\times10^{-7})$  & $c_3(\times10^{-10})$ \\
\hline
0     & 1.095 & 0.145 & 7.510     & $-$11.220 \\
15.99 & 0.445 & 2.217 & $-$30.583 & 9.635 \\
32.79 & 0.313 & 1.535 & $-$22.750 & 8.029 \\
\hline
\end{tabular} 
\end{table}

We compared our calculation results with other experimental data for cement paste and concrete.
Ukrainczyk {\it et al}.~\cite{Ukrainczyk} conducted experiments with calcium aluminate cement with w/c = 0.3, 0.4, finding that its thermal conductivity increases with increasing temperature from 293 K to 353 K due to loss of water from jennite at $343\sim363$ K.
At this temperature, the lattice constant $c$ of jennite crystal was found to shrink and simultaneously the jennite transformed from \ce{Ca9[Si6O_{18}(OH)6]}$\cdot$8\ce{H2O} to \ce{Ca9[Si6O_{16}(OH)2](OH)8}$\cdot$2\ce{H2O}, called metajennite~\cite{Yu}.
Xianzhi {\it et al}.~\cite{Xianzhi} demonstrated that the thermal conductivity of cement paste decreased like 0.82, 0.78, 0.60 W/m$\cdot$K at 333, 363 and 393 K.
Wang {\it et al}.~\cite{Wang} reported that the fly ash concrete had decreasing thermal conductivity from 1.69 to 0.95 W/m$\cdot$K as increasing temperature from 293 K to 793 K.
As they pointed out, the change of microstructure was found to start at 573 K from SEM image observation, and Yu~\cite{Yu} stated that jennite lost the interlayer water molecules and thus basic layer structure upon heating to 623 K.
It can be said that our calculation results are similar to these experimental data.
On the contrary, some irregular changing tendency was observed from experiment for different kinds of concrete.
For example, Gencel {\it et al}.~\cite{Gencel} found that the thermal conductivity of concrete, made using vermiculite as dominant component, decreased from 293 K to 1193 K and again increased to 1373 K.
Khaliq and Kodur~\cite{Khaliq} studied the temperature effect on the thermal properties for self-consolidating concrete, finding that its thermal conductivity decreased up to 693 K, marginally increased to 793 K, and again decreased to 1093 K.
For aluminous refractory concrete, Santos~\cite{Santos} reported the increasing tendency from 300 K to 333 K, significantly decreasing to 793 K, and again gradual increasing to 1273 K.
For explanation of such complicated behavior, the moisture loss by evaporation of free and pore water with the increase of temperature was considered as the main cause, which is different from our calculation models with the pore filled with air.

\section{\label{sec_con}Conclusions}
In this work, we have investigated the thermal conductivity of porous jennite by means of molecular dynamics method, aiming at clarifying the thermal behavior of cement hydration product C$-$S$-$H gel.
The porous jennite models with different porosities have been created by removing atoms within spherical region with a radius corresponding to the porosity, using $3\times4\times3$ supercell model for perfect jennite.
We have employed two different MD methods, namely equilibrium Green-Kubo and non-equilibrium M\"{u}ller-Plathe methods, to calculate the thermal conductivity of jennite along the crystallographic [100], [010] and [001] directions.
We found that the thermal conductivity of perfect jennite calculated by GK method using the supercell model were almost identical to those by MP method, exhibiting the crystallographic anisotropy like $K_2>K_1>K_3$.
The non-equilibrium MP method has been mostly applied in this work to investigate the porosity and temperature dependences of thermal conductivity of porous jennite, using the simulation boxes extended the porous jennite model 8 times along the heat flux direction and divided into 20 plane slices.
At room temperature of 300 K, the principal and volumetric thermal conductivities of bulk jennite were determined to be $K_1=1.100\pm0.025$, $K_2=1.354\pm0.034$, $K_3=0.967\pm0.020$ and $K_\text{v}=1.141\pm0.026$ W/m$\cdot$K in reasonable agreement with the experimental values for C$-$S$-$H.
We have found the decreasing tendency of effective thermal conductivity of porous jennite as increasing the porosity, which was confirmed to follow the empirical coherent potential or self-consistent model proposed for porous materials among other kinds of models.
Finally we have investigated the temperature dependence of thermal conductivity of porous jennite with the porosities of 0, 15.99 and 32.79\%, finding its gradual increase from 240 K to 560 K and decrease from 560 K to 1100 K.
By comparing with our own measurements and available experimental data for cement paste and concrete, the calculated values were said to be high reasonable and reliable.
Therefore, we believe this work will be helpful for a design of high thermal insulating materials with cement and concrete for energy efficient buildings.

\section*{\label{ack}Acknowledgments}
This work belongs to the state research project ``Design of Innovative Functional Materials for Energy and Environmental Application'' (No. 2016-20), supported partially by the State Committee of Science and Technology, Democratic People's Republic of Korea.
The simulations have been performed on the HP Blade System C7000 (HP BL460c) that is managed by Faculty of Materials Science, Kim Il Sung University.

\section*{\label{note}Notes}
Declarations of interest: none.

\bibliographystyle{elsarticle-num-names}
\bibliography{Reference}

\end{document}